\documentclass[fleqn,usenatbib]{mnras}
\usepackage{newtxtext,newtxmath}
\usepackage[T1]{fontenc}

\DeclareRobustCommand{\VAN}[3]{#2}
\let\VANthebibliography\thebibliography
\def\thebibliography{\DeclareRobustCommand{\VAN}[3]{##3}\VANthebibliography}

\usepackage{graphicx}	
\usepackage{amsmath}	
\usepackage{hyperref}
\usepackage{gensymb}

\hypersetup{
    colorlinks=true,                
    breaklinks=true,                
    urlcolor= blue,                  
    linkcolor= blue,                
    citecolor= blue                 
    }

\title[Jet–red giant interactions as a source of extragalactic neutrinos: Insights from KM3-230213A]{Jet–red giant interactions as a source of extragalactic neutrinos: Insights from KM3-230213A}

\author[G. Fichet de Clairfontaine et al.]{
G. Fichet de Clairfontaine,$^{1}$\thanks{E-mail: gaetan.fichet@uv.es}
M. Perucho,$^{1,2}$
and J.M. Martí$^{1,2}$
\\
$^{1}$Departament d’Astronomia i Astrof\'isica, Universitat de Val\`encia, C/ Dr. Moliner, 50, E-46100 Burjassot, Val\`encia, Spain\\
$^{2}$Observatori Astron\`omic, Universitat de Val\`encia, C/ Catedr\`atic Jos\'e Beltr\'an 2, E-46980 Paterna, Val\`encia, Spain\\
}

\date{Accepted for publication in MNRAS.}

\pubyear{\the\year{}}

\begin{document}
\label{firstpage}
\pagerange{\pageref{firstpage}--\pageref{lastpage}}
\maketitle

\begin{abstract}
The production sites of high-energy astrophysical neutrinos remain uncertain, though growing evidence suggests a connection to relativistic jets in active galactic nuclei (AGN). We present a detailed analysis of the recent PeV neutrino event KM3$-$230213A reported by the KM3NeT collaboration, aiming to constrain the physical conditions of its source. Assuming proton acceleration at shocks, we derive the properties of the proton distribution and the energetics required to explain the neutrino emission. Using contemporaneous multiwavelength observations of three AGN flaring candidates within the error region, we examine the plausibility of each of them as the possible counterpart. Our results favor PMN\,J0606$-$0724, which exhibits a prominent radio flare coincident with the neutrino arrival. In this framework, the red-giant interaction remains the key driver of baryon injection and shock acceleration, while the dominant external photon field sets the neutrino energy scale: photospheric photons from the red giant yield $\sim1-10~\rm{PeV}$ neutrinos, whereas the $\sim 220~\rm{PeV}$ event KM3$-$230213A is more naturally produced through interactions with colder infrared photons from the dusty torus.
\end{abstract}


\section{Introduction}
\label{sec: introduction}

The detection of high-energy neutrinos provides a unique window into the most energetic and powerful events in the Universe. Recent advances in neutrino astronomy, particularly by the KM3NeT collaboration \citep{Km3net_2025}, could significantly improve our understanding of these phenomena. Extragalactic neutrinos are believed to originate from various sources, including active galactic nuclei (AGN), gamma-ray bursts (GRBs), and supernovae. Among these, AGN jets aligned with the line of sight (blazars) are considered promising candidates because they represent a high-energy scenario plausibly conducive to neutrino production \citep{Murase_2023}. Stacking analyses of \textit{Fermi} 2LAC blazars \citep{Aartsen_2017}, radio-bright AGN \citep{Abbasi_2023}, and hard-X-ray-selected AGN \citep{Abbasi_2025} show that the blazar population can contribute at most a sub-dominant ($\lesssim10$–20 \%) fraction of the diffuse TeV–PeV neutrino flux. Nevertheless, blazar jets stay among the most likely objects that could be responsible for the \emph{individual} neutrino emission event discussed here, even though the class as a whole cannot explain the bulk of the diffuse flux. That being said, theoretical models suggest that interactions between relativistic protons in AGN jets and ambient photon fields can lead to photopion production, resulting in the emission of high-energy neutrinos in the PeV range.
Blazars draw more attention because of the presence of Doppler boosting that enhances the observed emission. Different theoretical works have shown that the neutrino production channel can be significant in AGN jets, where intense photon fields (synchrotron self-Compton, external fields) and relativistic protons could coexist \citep{Mannheim_1993, Murase_2014, Murase_2022}. Therefore,  neutrino detection can be crucial to set constraints on the proton energy and the conditions at the emission regions in these sources. 

Recent studies point out the underlying link between neutrino detection and the presence of multi-wavelength flaring activity. This observational tool was used intensively in the radio \citep{Kouch_2024, Suray_2024, Allakhverdyan_2024}, X-ray \citep{Stathopoulos_2022} and $\gamma$-ray \citep{Peng_2017, Yoshida_2023} bands.  Although no definitive associations between individual high-energy events and their cosmic sources have been firmly established, several potential connections with AGN have been suggested at varying statistical significance. For instance, the blazar TXS\,0506$+$56 has been linked to high-energy neutrinos detected by the IceCube neutrino observatory in multiple studies \citep{IceCube_2018a, IceCube_2018b}. Other methods to explore such correlations involve the use of time-integrated neutrino data over specific periods. One example is the strongest observational evidence to date for neutrino emission in the direction of the Seyfert galaxy, NGC\,1068 \citep{IceCube_2022}. Additionally, stacking analyses of populations of AGN sources have been employed to investigate potential neutrino emissions from these objects \citep[e.g.][]{IceCube2017_2LAC,Padovani_extreme_blazars:2016, Aartsen_2017, IceCube_AGNcores:2021,Plavin_2021,IceCat-1:2023}. 

Moreover, theoretical work has related that neutrinos to large electromagnetic flares \citep{Gao_2019, Rodrigues_2019, Rodriguez_2021}. A notable example is the work by \citet{Mastichiadis_2021}, who pointed out that by-products of photopion interactions are expected to emit at the X-ray band, especially during flaring events.

A $220~\rm{PeV}$ neutrino detection on the $23^{\rm rd}$ of February $2023$ was recently reported by the KM3NeT collaboration. This observation, named KM3-230213A, marks a milestone in neutrino astronomy as it is the first time that a PeV neutrino, predicted in the context of photopion interactions \citep{Dermer_2014}, is observed. Blazar jets are most likely responsible for the neutrino emission. Indeed, seventeen blazar candidates were found within $3\degree$ around these neutrino event coordinates \citep{km3net_2025b}. The steady isotropic flux that would produce the discussed event is
\begin{equation} \label{eq:flux}
    E_{\rm \nu, obs}^2 \Phi_{\rm \nu, obs} = 5.8 \times 10^{-8}~\rm{GeV}\cdot\rm{cm}^{-2}\cdot\rm{s}^{-1}\cdot\rm{str}^{-1}\,.
\end{equation}
This can be used to provide critical information about the proton luminosity and set constraints on the energetics required by a given source to produce such an extremely energetic neutrino. Analyzing the detected neutrino’s energy and flux allows us to estimate the proton energy distribution, luminosity, and energy density. This is crucial to set constraints on the emitting region. In fact, the recent cosmic–ray analysis by \citet{Das_2025} places a stringent ``blazar horizon’’ at roughly $z \simeq 1$: above this redshift, a single PeV–EeV neutrino detected by KM3NeT would require proton luminosities that overshoot the cosmic-ray energy budget. Conversely, \cite{Neronov_2025} argues that KM3-230213A may stem from a year-long transient, re-opening the possibility of more extended proton-cooling regions in the jet. In this work, we propose a scenario that favors a transient, extreme event as a potential origin of the stringent flux detected by the KM3NET collaboration, as it was proposed in the past for TXS\,0506$+$056 \citep{Murase_2018}. In fact, our results show a departure from equipartition between the non-thermal protons and the magnetic field, which is easier to reconcile with a transient event as caused by a sudden increased contribution of protons to the radiative output produced at shocks in different scenarios (e.g., the injection of energetic perturbations at the jet-forming region).  \\

This paper is organized as follows. In Sect.~\ref{sec: equations}, we develop the derivation of the proton distribution, luminosity, and energy density. In Sect.~\ref{sec: sources}, we discuss each of the identified flaring sources by the KM3NeT collaboration in the frame of our modeling. In Sect.~\ref{sec:scenario}, we discuss a plausible scenario for PeV neutrino emission, and apply it to the KM3$-$230213A event. Finally, in Sect.~\ref{sec: conclusion}, we summarize our findings. Throughout the paper, quantities directly related to the KM3$-$230213A event are labeled “obs”, while all other quantities are calculated in the context of an assumed flat $\Lambda$CDM cosmology with ${H}_{\rm 0} = 69.6~\rm{km}\cdot\rm{s}^{-1}\cdot\rm{Mpc}^{-1}$, $\Omega_{\rm 0} = 0.29$, and $\Omega_{\Lambda} = 0.71$. 

\section{Derivation of the proton distribution}
\label{sec: equations}

The energy of the neutrino observed by the KM3NET collaboration has been estimated to be $E_{\rm \nu, obs} \geq 220~\rm{PeV}$ \citep{Km3net_2025}. In fact, the measured muon energy provides a lower bound on the incoming neutrino energy. Based on the estimated muon energy and its uncertainty, simulations of the ARCA detector indicate that the median neutrino energy producing such muons is $220~\rm{PeV}$, an energy that we will consider in the rest of this study. If the neutrino arose from proton–photon ($p\gamma$) collisions —specifically via the $\Delta^+$ ($1232~\rm{MeV}$) resonance— then there is a minimum proton energy required to excite that resonance.  In the proton’s rest frame, the $\Delta^+$ channel produces charged pions ($\pi^\pm$) carrying on average about $20\%$ of the proton’s energy when the target photons are relatively low‐energy \citep[and references therein]{Mannheim_1995}.  Consequently, the requirement to reach the $\Delta^+$ threshold and the fact that each pion only inherits $\sim~1/5$ of the proton’s energy together set an upper bound on the proton Lorentz factor in any $p\gamma$–driven neutrino source. This ratio, often called inelasticity, is $\kappa_{\rm p \gamma} (\simeq 0.2)$. This means that, on average, $20\%$ of the proton energy goes into the products of pion decays. Furthermore, for each proton-meson interaction, $3$ different neutrinos can be produced in addition to a lepton. Then, $E_{\rm \nu} \sim (\kappa_{\rm p \gamma} / 4) \, E_{\rm p}$. This leads to the commonly used relation
\begin{equation}
    E_{\rm p, obs} \simeq 20 E_{\rm \nu, obs},
\label{eq: E_p_obs}
\end{equation}
where $E_{\rm p, obs}$ is the proton energy that leads to the production of the neutrino at the corresponding energy. The proton Lorentz factor is $\gamma_{\rm p} = E_{\rm p} / (m_{\rm p} c^2)$. The neutrino flux derived has been estimated to be 
\begin{equation}
    \Phi_{\rm \nu, obs} = 1.87 \times 10^{-24}~\rm{erg}^{-1}\cdot\rm{cm}^{-2}\cdot \rm{s}^{-1} \cdot \rm{str}^{-1},
\end{equation}
which is obtained by dividing Eq.~\ref{eq:flux} by $E_{\rm \nu, obs}^2$.
From the previous expression, we can deduce the hypothetical neutrino spectrum as observed on Earth
\begin{equation}
    \Phi_\nu \left(E_{\nu}\right) = \Phi_{\rm \nu, obs} \left(\dfrac{E_\nu}{E_{\rm \nu, obs}}\right)^{-2} \,.
\end{equation}
Considering a typical power-law distribution for the relativistic protons accelerated by the Fermi process on shocks in the jet \citep{Lemoine_2019}, i.e.,
\begin{equation}
    \dfrac{\mathrm{d}N_{\rm p}}{\mathrm{d}E_{\rm p}} = K_{\rm p} E_{\rm p}^{-2} \,,
    \label{eq:distrprotons}
\end{equation}
and assuming that the distribution spreads between the boundary values $E_{\rm p,min} = m_{\rm p} c^2$ and $E_{\rm p,max} = E_{\rm p,obs}$, which are taken to be the proton energy at rest and the proton energy associated to the observation. Although the proton distribution formally extends down to rest mass, only protons above the photopion threshold can produce neutrinos. For a target photon field in the ultraviolet, this corresponds to $E_{\nu,\min}\simeq 100~\mathrm{TeV}$ \citep{Murase_2014}. We therefore integrate the neutrino flux between this lower limit and $E_{\nu,\max}=0.05\,E_{\rm p,max}$ to obtain the neutrino luminosity in the lab frame,
\begin{equation}
    L_{\rm \nu} \simeq \Omega_{\rm obs} \ D_{\rm L}^2 \int_{E_{\rm \nu, min}}^{E_{\rm \nu, max}} \mathrm{d}E~E  \cdot \Phi_{\rm \nu}\left(E\right) \,,
\label{eq: L_p}
\end{equation}
with $\Omega_{\rm obs} \simeq \pi (1 + z) / \delta_{\rm d}^2$ the solid angle of a narrow beamed jet ($\theta_{\rm obs} \sim \delta_{\rm d}$, with $\delta_{\rm d}$ the jet Doppler factor), and $D_{\rm L}$ the luminosity distance of the source. Following \cite{Murase_2014, Murase_2018}, one can directly relate the neutrino luminosity to the proton luminosity in the context of the $p\gamma$ process, 
\begin{equation}
    L_{\rm p} = \dfrac{8}{3} \dfrac{E_\nu}{E_{\rm p}} \dfrac{1}{f_{\rm p \gamma}} L_\nu  \,,
    \label{eq: Murase}
\end{equation}
$L_{\rm p}$ being the proton luminosity. Here, $f_{\rm p\gamma} \sim 10^{-3}$ denotes the photomeson production efficiency, which quantifies the fraction of high-energy protons that interact with ambient photons to produce pions (and subsequently neutrinos). This value is characteristic of blazar environments where the photon target density is relatively low, such as in the jet regions of radio-loud AGN \citep{Dermer_2014}. The factor $3/8$ corresponds to the average energy fraction and multiplicity in photopion production processes leading to neutrinos. As discussed in \cite{Murase_2018}, such low interaction efficiencies are consistent with the estimated transparency of these sources to gamma rays, and set an important constraint on the expected neutrino flux. It should be noted that Eq.~\ref{eq: Murase} relates the neutrino luminosity to the luminosity of those protons above the photopion threshold, i.e., the participating proton population (with $E_{\rm p, thr} = 20 E_{\rm \nu, min}$). For a power-law spectrum with index $2$, the fraction of the total proton luminosity that participates is, 
\begin{equation}
    F_{\rm thr} = \ln \left(E_{\rm p,\max}/E_{\rm p, thr}\right) / \ln\left(E_{\rm p,\max}/E_{ \rm p,\min}\right)\,,
\end{equation}
so that the bolometric proton luminosity can be recovered as $L_{\rm p,total} = L_{\rm p} / F_{\rm thr}$. \\

We can then make a simplification and consider that protons are accelerated within a cylinder of radius $R$. Then, we can derive the proton energy density as follows,

\begin{equation}
    u_{\rm p} = \dfrac{L_{\rm p, total}}{\pi R^2 c}\,.
\end{equation}

It is useful to compare this quantity with the magnetic energy density $u_{\rm B} = B^2 / 8 \pi$ at the acceleration site and estimate the ratio $\chi = u_{\rm p} / u_{\rm B}$ and its possible deviation from equipartition: $\chi \sim 1$. 

The number density of the relativistic protons can be obtained by solving for $N_{\rm p}$ (see Eq.~\ref{eq:distrprotons}) in the following expression,
\begin{equation}
    N_{\rm p} = \dfrac{K_{\rm p}}{1 - p} \left[E_{\rm p,max}^{1-p} - E_{\rm p, min}^{1-p}\right] \,,
\end{equation}
where $K_{\rm p} = u_{\rm p} / \int E_{\rm p}^{1-p}\text{d}E_{\rm p}$ is the normalization factor of the proton distribution, and the index $p = 2$ in our case. Following up on the hypothesis that protons are accelerated at shocks, we can derive the typical acceleration timescale,
\begin{equation}
    \tau_{\rm acc} = \dfrac{1}{\eta} \dfrac{m_{\rm p} c \gamma_{\rm p}}{e B} \,,
\end{equation}
where $\eta = 0.1$ is a free parameter that characterizes the shock acceleration efficiency \citep[chosen here to be consistent with what derived from particle-in-cell, PIC, simulations and previous modeling][]{Sironi_2011, Cerruti_2015}.
In AGN jets, the fastest cooling timescale for protons is generally the advection timescale $\tau_{\rm ad} = R/c$ \citep{Zech_2017}, compared to the radiative cooling timescale. Here, $R$ stands for the radius of the previously mentioned cylinder, assuming that the dominant escape route is transverse due to the gyro movement of the particles. In such a scenario, the gyro radius $R$ of the proton increases during the acceleration time $\tau_{\rm ad}$, and we assume that the final $R$ is such that $R \leq R_{\rm jet}$ where $R_{\rm jet}$ is the jet radius. In other words, we consider that protons are still confined within the jet at the time of the $p\gamma$ interaction. Therefore, a maximal proton Lorentz factor can be found by equating the acceleration to the advection timescale,
\begin{equation}
    \gamma_{\rm p, max} = \eta \dfrac{e B R}{m_{\rm p}c^2} \,.
\end{equation}

It should be noted that in a medium where the photon field is dense enough, $p\gamma$ losses may become faster than the light-crossing time, which will reduce the value of $\gamma_{\rm p, max}$ \citep{Murase_2014}. As we infer the proton energy from the neutrino observation, we will assume, for simplicity, that the confinement limit holds.

\section{Discussion on the identified sources}
\label{sec: sources}

\begin{figure}
    \centering
    \includegraphics[width=\linewidth]{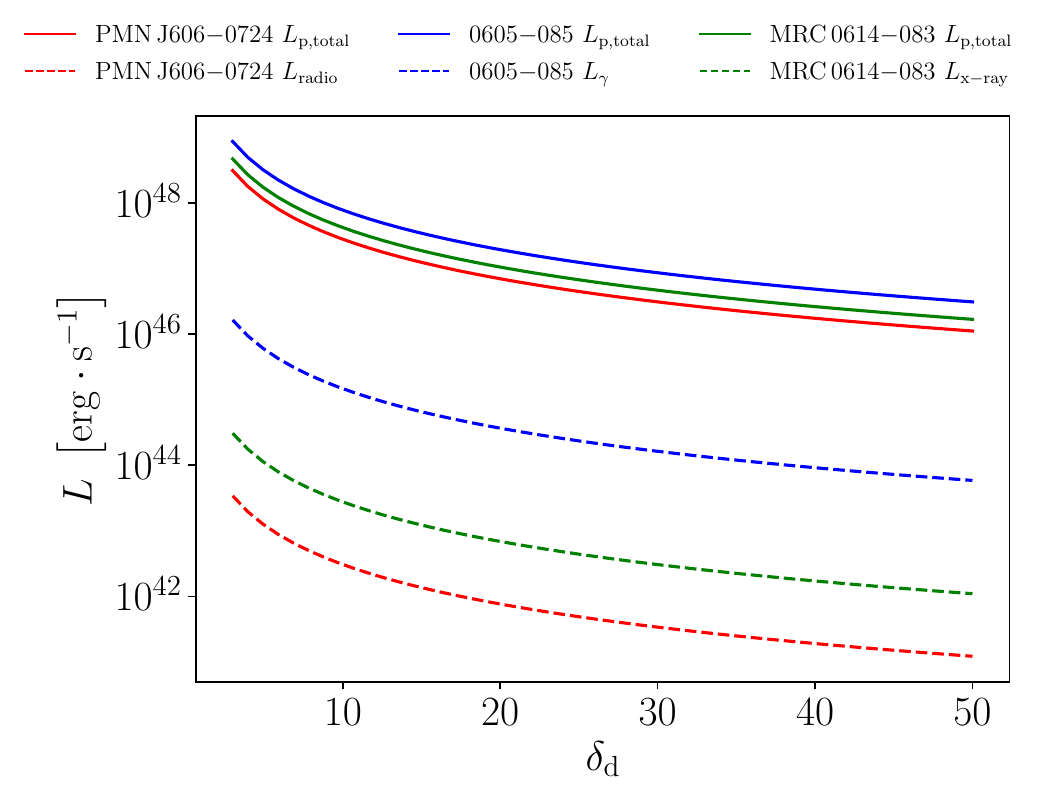}
    \caption{Total proton luminosity as observed in the lab-frame as a function of the Doppler factor. Three different redshifts are used for the three sources of interest (see text). We also show the radio luminosity (OVRO, at $15~\rm{GHz}$) for PMN\,J606$-$0724, the $\gamma$-ray luminosity at $1~\rm{GeV}$ (\emph{Fermi}-LAT) for 0605$-$085, and the X-ray luminosity integrated between $0.2 - 2.3~\rm{keV}$ (eROSITA) for MRC\,0614$-$083.}
    \label{fig:L_plot}
\end{figure}

In a follow-up study, \cite{km3net_2025b} identifies a set of $17$ blazar candidates that lie within the $99\%$ confidence region around the estimated emitting location, which implies an angular radius of $3\degree$. This represents an increase of five sources with respect to \cite{Km3net_2025}. The authors discuss the extensive correlation method and the data collection performed for each of the $17$ sources. All the details on the data collection, light curves, etc., can thus be found in \cite{km3net_2025b}. Among the $17$ sources, the authors highlight $3$ of them that undergo flaring activity around the neutrino detection. We sum up their findings about these three sources for the reader as 
\begin{itemize}
    \item 0605$-$085 (RFC\,J0607$-$0834): a \emph{Fermi}-LAT \citep{Fermi_2009} source that shows gamma-ray variability ($\tau_{\rm flare} \sim 2~\rm{years}$, in the falling phase during the neutrino arrival time), with an energy flux of $\sim 5 \times 10^{-8}~\rm{ph}\cdot\rm{cm}^{-2}\cdot\rm{s}^{-1}$ integrated between $0.1$ - $1000~\rm{GeV}$. This source is also the brightest radio source among the sample, with a flux of $\sim 2.240~\rm{Jy}$ observed at the GHz band. The source is at an estimated redshift of $z = 0.87$ \citep{Shaw_2012}.
    \item PMN\,J0606$-$0724 (RFC\,J0606$-$724): this radio source, observed by the Owens Valley Radio Observatory (OVRO, \cite{Ovro_2011}) and the RATAN instruments \citep{Ratan_1992} at $15~\rm{GHz}$, displays an active flaring history, with a radio flare ($\tau_{\rm flare} \sim 1~\rm{year}$) that peaked precisely only five days after the neutrino detection with a flux of $\sim 0.7~\rm{Jy}$. Its estimated redshift is $z = 1.227$ \citep{Healey_2008}.
    \item MRC\,0614$-$083 (NVSS\,J061703$-$082225): this source displayed a sparse X-ray activity, observed before and after the neutrino arrival time. The light curve at this band shows flaring activity ($\tau_{\rm flare} \sim 3~\rm{years}$), but the sparsity of the data doesn't allow us to assess if a peak or a plateau is reached at the time of the neutrino detection. Nevertheless, an energy flux of $\sim 10^{-12}~\rm{erg}\cdot\rm{cm}^{-2}\cdot\rm{s}^{-1}$ integrated between $0.2 - 2.3~\rm{keV}$ has been reported \citep[eROSITA;][]{erosita_2012}. The redshift of this source is unknown.
\end{itemize}

\begin{figure}
    \centering
    \includegraphics[width=\linewidth]{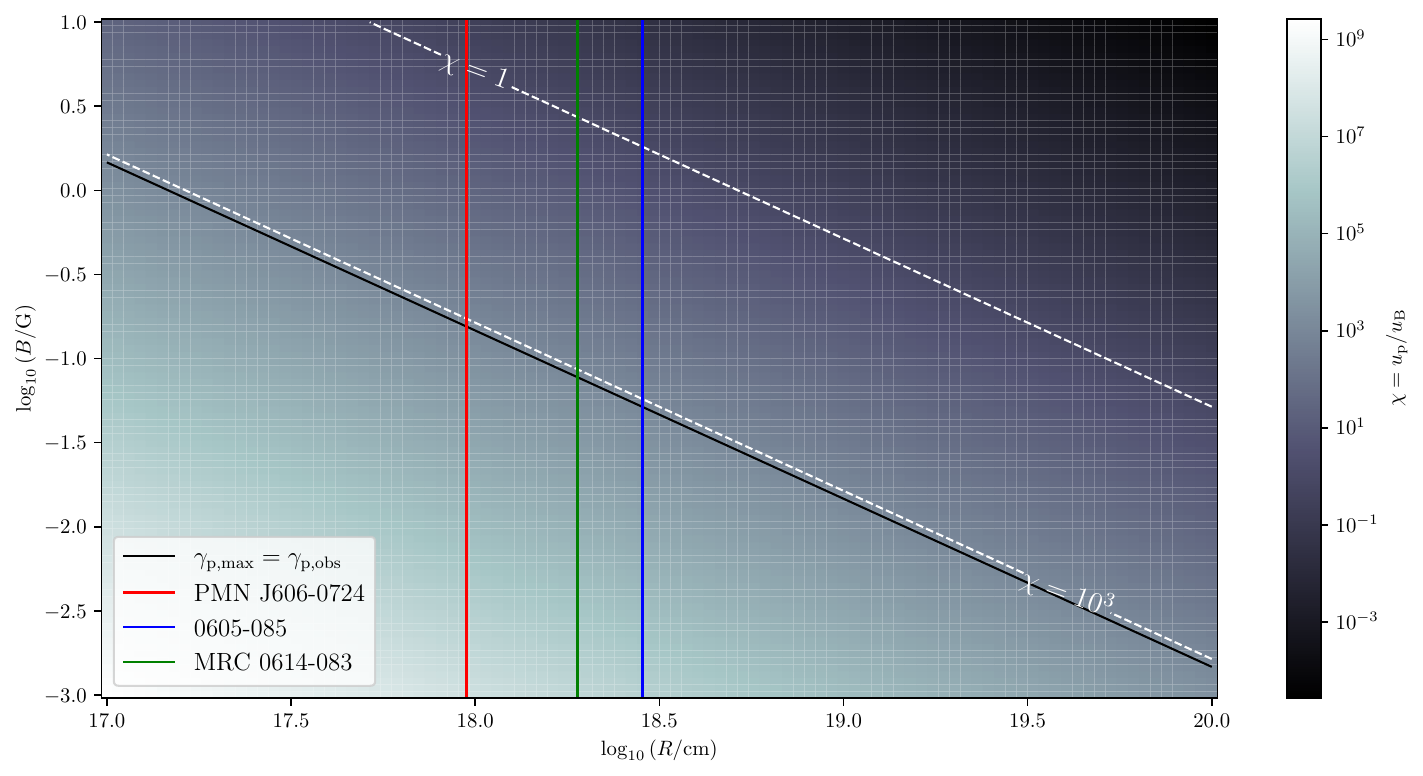}
    \caption{Ratio of proton energy density $u_{\rm p}$ over the magnetic one $u_{\rm b}$, as a function of the magnetic field strength $B$ and the size of the emission region $R$, for a proton luminosity of $L_{\rm p} = 10^{47}~\rm{erg}\cdot\rm{s}^{-1}$. White contour lines represent ratio values of $1$ (equipartition) and $10^3$. Vertical colored lines indicate estimates of the emitting region sizes of the three selected sources.}
    \label{fig:equipartition_array}
\end{figure}

In Fig.~\ref{fig:L_plot} we show the derived proton luminosity from Eq.~\ref{eq: L_p}, using the redshift of each source and varying the Doppler factor. For the calculations, we have assumed $z=1$ for MRC\,0614$-$083 (see \cite{Das_2025}). The plot shows that the luminosity lies between $L_{\rm p} \simeq 10^{46}$ and $10^{48}~\rm{erg}\cdot\rm{s}^{-1}$ for Doppler factors ranging from $3$ to $50$. Using this information, we can estimate $\chi$ in the different sources, following the equations given in the previous Section. We show this parameter in Fig.~\ref{fig:equipartition_array} as a function of the size of the emitting region and magnetic field for the given proton luminosity. 

Then, for each source, we estimate an upper limit of the size of the emitting region by using the flare timescale: $R \sim \tau_{\rm flare} \cdot  c$. We use the flaring time from $\gamma$-ray, radio, and X-ray light curves, respectively, as given in \cite{km3net_2025b}. The vertical colored lines indicate the resulting value for each source in Fig.~\ref{fig:equipartition_array}. In all cases, the derived size of the emitting region is $R \leq 1~\rm{pc}$. 

Taking the advection timescale as the cooling timescale and assuming $\gamma_{\rm p, max} = \gamma_{\rm p, obs}$, we can estimate the value of $\chi$. The result is shown by a solid black line in Fig.~\ref{fig:equipartition_array}. Interestingly, under the above assumptions, the region responsible for the neutrino emission seems to be far off equipartition, with $\chi \simeq 10^3$, which corresponds to a high proton luminosity of $L_{\rm p} \simeq 10^{47}~\rm{erg}\cdot\rm{s}^{-1}$.

Before proceeding, let us note that, with this estimate and the sizes of the emitting regions $R$ (as derived in the previous paragraph), we can derive a value for the magnetic field intensity $B$ in the observer's frame from the condition $\gamma_{\rm p,max} = \gamma_{\rm p,obs}$ for the different emitting region sizes (intersections of the solid black line with the vertical colored ones in Fig.~\ref{fig:equipartition_array}), which turns out to be of the order of $\sim 0.1~\rm{G}$ in the three cases. Now, coming back to the value of $\chi$, the significant departure from equipartition would point towards a transient, dramatic event taking place in the jet. Such an event would explain both the photon and neutrino flares. We discuss the plausible scenarios that could take place in the next Section.

In the case of 0605$-$085, the observed $\gamma$-ray luminosity is $L_\gamma \leq 10^{46}~\mathrm{erg}~\rm{s}^{-1}$ at $1~\rm{GeV}$. However, the significant delay between the neutrino arrival time and the $\gamma$-ray flaring peak activity, together with the absence of noticeable radio variability during the same period, both indicate that the observed neutrino is unlikely to have originated from 0605$-$085.

In the case of the radio flaring source PMN\,J0606$-$0724, the large deviation from equipartition that we have derived can be associated with a transitory event related to the radio flare. In fact, we can speculate that an increase in the proton energy density is related to an event of copious proton acceleration in the jet, with the resulting increase in their number density and subsequent acceleration. It should be noted that no significant $\gamma$-ray emission from the source was detected during the neutrino event, and only a high upper-limit resulting from a $6$-month integration was derived \citep{km3net_2025b}. The derived high proton luminosity, $L_{\rm p} \sim 10^{47}~\rm{erg}\cdot\rm{s}^{-1}$, aligns with the source’s considerable distance \citep{Das_2025}. At such high redshifts, only sources with intrinsically powerful emissions are detectable, as less luminous emissions, including $\gamma$-rays, may be too faint to observe due to cosmological dimming. Therefore, the absence of detected $\gamma$-ray emission could result from either an intrinsic lack of significant $\gamma$-ray production or the attenuation of such emissions over vast cosmological distances.

MRC\,0614$-$083 has an unknown redshift, which challenges any predictions for this source. If we assume its redshift to be $\sim 1$, the required proton budget is significantly increased compared to the magnetic energy. The source is characterized by a particular X-ray activity. At the assumed redshift ($z = 1$), the peak X-ray flux is $L_{\rm peak, \, X-ray} \lesssim 10^{44}~\rm{erg}\cdot\rm{s}^{-1}$, or $\sim 10^{-3} \, L_{\rm \nu, obs}$. This value is small enough to give an X-ray-to-neutrino luminosity ratio still much smaller than $1$, when integrated over the whole X-ray band. According to recent theoretical studies \citep{Petropoulou_2020, Mastichiadis_2021, Fichet_2023}, values of that ratio close to $1$ would represent evidence of emission from hadronic processes contributing to the X-ray band. Hence, although the sparsity of the X-ray observations during the neutrino arrival time makes it difficult to extract definite conclusions, the small value of this ratio in the case of MRC\,0614$-$083 would indicate that the X-ray activity is unlikely related to hadronic processes (e.g., pair production or pion decay) leading to a significant neutrino activity.

In summary, among all flaring candidates flagged by the KM3NeT Collaboration, the radio-bright blazar PMN\,J0606$-$0724 emerges as the most plausible counterpart of KM3$-$230213A. The radio flare, whose peak follows the neutrino arrival by only $\sim 5~\rm{days}$, is a strong circumstantial clue, although it cannot, by itself, prove the association. Our analysis implies a large departure from equipartition and a proton luminosity $L_{\rm p}\simeq 10^{47}~\rm{erg}\cdot\rm{s}^{-1}$, values that are fully consistent with the year-long, high-power transient scenario proposed by \cite{Neronov_2025}. In the next section, we sketch one concrete mechanism based on the rapid acceleration of external matter, followed by efficient shock or shear acceleration, that can simultaneously trigger the observed radio outburst and supply the required proton energy budget to produce the PeV neutrino.

\section{Model}
\label{sec:scenario}

Blazars are expected to contribute marginally to the diffuse neutrino flux at very-high energy $<\rm{PeV}$. Nonetheless, $17$ known blazar sources have been located within $3^\circ$ of the estimated position of the event KM3$-$230213A. Furthermore, blazars are expected to be an ideal site of neutrino production due to proton acceleration through shocks and $p\gamma$ interactions within a dense photon field. In the following section, we describe a potential scenario that could explain the emission of PeV neutrinos and potentially apply it to the KM3$-$230213A-like events. 

\subsection{A plausible scenario: shock-red-giant interaction}

We consider a jet composition initially dominated by leptons, as expected if the jet is formed by the Blandford--Znajeck (BZ) mechanism \citep{Blandford_1977}, and as numerical simulations \citep[e.g.,][]{McKinney_2012} and modeling of observations \citep{EHT_2019} seem to indicate. Because of this initial composition, any neutrino production process taking place in AGN jets at the sub-pc to pc scales would require the entrainment and acceleration of protons to the necessary energies. This alteration of the jet composition demands either an external source or a modification in the source's accretion/ejection conditions.\\

An appealing scenario to explain the transient proton loading required for the observed neutrino production involves jet–star interactions, particularly with red-giant (RG) stars. During its propagation, the relativistic jet can collide with dense stellar atmospheres or winds from RG stars present in the circumnuclear region, as detailed by several works \citep[e.g.][]{Barkov_2010, Bosch_2012, Fichet_2025}. Such interactions can effectively entrain stellar-wind material into the jet, substantially enriching it with protons in localized regions. Once protons are entrained, acceleration mechanisms—predominantly shocks or turbulent mixing layers produced by the interaction with the stellar obstacle \citep{Araudo_2013, de_la_Cita_2016, Perucho_2017, 2020MNRAS.494L..22P}—can efficiently energize particles. However, the number density of resulting relativistic protons is crucial to determine whether significant $p\gamma$ interactions occur. Red giants are especially efficient compared to main-sequence (MS) stars: while a typical MS star loses mass at a rate of $\dot{M}_{\rm MS} \sim 10^{-14}- 10^{-12}~M\odot\cdot\rm{yr}^{-1}$ \citep[and references therein]{Lanz_1992}, red giants can reach $\dot{M}_{\rm RG} \sim 10^{-8}-10^{-5}~M\odot\cdot\rm{yr}^{-1}$ \citep{Reimers_1977}. Their bow shocks are also substantially larger, with stand-off radii $R_{\mathrm{bs}} \sim 10^{15}-10^{16}~\rm{cm}$ compared to $R_{\mathrm{bs}} \lesssim 10^{13}~\rm{cm}$ for MS stars, enhancing both the volume of shocked plasma and the efficiency of particle injection into the jet. Such interactions sustain higher proton loading and more effective pre-acceleration, setting the stage for distinct radiative outcomes depending on the jet’s dynamical state. One can therefore expect steady radio emission from the extended jet, but only a rather low neutrino output. In the quiescent case, the baryons injected by the red giant are not sufficient to produce a detectable $p\gamma$ flux, because the target photon field is only modestly boosted into the jet frame and the interaction efficiency remains very low. However, if a moving shock or ejecta is launched from the jet base while the RG is still crossing the jet, the baryon density and the local photon field are suddenly enhanced, and the situation can dramatically change.

First, the perturbation would propagate and locally accelerate electrons, triggering a slow but steady radio rise \citep[e.g.,][]{Fromm_2011,Fromm_2013,Fromm_2016}. Past numerical works have shown how such kind of perturbations can explain the radio flares commonly observed in radio galaxies \citep{Marscher_1985, Daly_1988,Gomez_1997,2009ApJ...696.1142M,Fromm_2016,Fichet_2021, Fichet_2022}. In contrast, if the shock passes through the RG, it will sweep the proton reservoir, significantly increasing the number density of relativistic protons, and further accelerating them during the interaction. Moreover, the dense photon field, necessary for significant $p\gamma$ neutrino production, is boosted in the jet frame. As protons collide with this photon field, they can undergo photopion ($p\gamma$) interactions and lead to neutrino production. 

A key observational signature of jet–red-giant interactions is that they can produce detectable neutrino emission without a bright, coincident GeV–TeV $\gamma$-ray counterpart. Neutral pions produced in $p\gamma$ interactions decay into $\gamma$-rays with a luminosity comparable to the neutrino channel, $L_{\gamma,\pi^0}/L_{\nu}\approx 4/3$. For our fiducial values ($L_{p}\sim10^{47}~\rm{erg}\cdot\rm{s}^{-1}$, $f_{p\gamma}\sim10^{-3}$) this corresponds to $L_{\gamma,\pi^0}\sim5\times10^{43}~\rm{erg}\cdot \rm{s}^{-1}$, or an observed flux of $F_{\gamma,\pi^0}\sim5\times10^{-15}~\rm{erg}\cdot\rm{cm}^{-2}\rm{s}^{-1}$ at $z\simeq1.2$ (corresponding to PMN\,J0606-0724). Even if all this power emerged inside the $0.1$-$100~\rm{GeV}$ band, it would remain $10^2$-$10^3$ below the detection threshold of \textit{Fermi}-LAT. In reality, the injected $\pi^0$ photons lie at $\gtrsim~\rm{PeV}$ energies and are immediately absorbed by pair production ($\gamma\gamma$) on the boosted stellar UV field, producing $e^\pm$ pairs that cool rapidly by synchrotron. This initiates an electromagnetic cascade that reprocesses most of the $\pi^0$ power into the MeV–GeV band rather than TeV-PeV. Additional $\gamma\gamma$ absorption on external photon fields (e.g., BLR or dusty torus) and on the extragalactic background light (EBL) further suppresses any surviving high-energy component. This is consistent with detailed modeling by \citet{Murase_2016, Reimer_2019}, which shows that in such environments, significant neutrino production does not necessarily imply a detectable $\gamma$-ray signal. We therefore conclude that the predicted MeV-GeV $\gamma$-ray flux associated is well below current observational sensitivities for sources at $z \sim 1$. However, if detectable, one will expect an associated MeV-GeV gamma-ray flare close to the neutrino arrival time according to our scenario.\\ 

One should mention that cloud or star ablation by jets has been invoked in the past to explain major $\gamma$-ray flares \citep{Zacharis_2017}, typically when the object enters the jet and is progressively disrupted. In our case, the RG is already embedded within the jet, leading to a quasi-steady interaction and being disturbed only during the interaction with the moving shock. While we cannot exclude the possibility of a $\gamma$-ray flare during the initial entry phase, its variability timescale would be much longer than that of the radio flare or neutrino outburst, given the RG’s extended crossing time, of the order of hundreds to thousands of years. \\

More precisely, typical RG emit thermal photons peaking at $\epsilon_{\star}\sim1.2~\rm{eV}$, for $T_\star \sim 5000~\rm{K}$ \citep{Piau_2011}. In the jet rest frame {(primed quantities)} these photons are Doppler–boosted by the jet bulk Lorentz factor $\Gamma \sim 10$, with respect to the AGN frame, so that $\epsilon^\prime_{\star} \simeq 2\Gamma\epsilon_{\star} \sim\ 24~\rm{eV}$, placing them in the near-ultraviolet. The stellar photon density in the jet frame is
\begin{equation}
    n^\prime_{\rm ph, \star} = \dfrac{2\Gamma L_{\star}}{4\pi\,R_{\rm s}^2\,c\,\epsilon_{\star}} \sim 10^8~\rm{cm}^{-3}\,,
\end{equation}
with $L_{\star}\sim10^3~L_\odot$ being the RG luminosity and $R_{\rm s}\sim10^{15}~\rm{cm}$ the stagnation radius (extension of the stellar photon field in the jet). These boosted stellar photons then dominate over the local synchrotron photon field as $p\gamma$ targets, without requiring an intrinsically extreme synchrotron brightness \citep{Araudo_2011}. The resulting $p\gamma$ {absorption depth} is $\tau_{\rm p\gamma} \sim n^\prime_{\rm ph, \star} \sigma_{\rm p\gamma} R_{\rm s} \sim 10^{-4}$ ($\sigma_{\rm p\gamma}$ being the $p\gamma$ cross-section), consistent with the hypotheses made in Sect.~\ref{sec: equations}. Crucially, this mechanism activates precisely when the moving shock passes through the dense photospheric shell of the RG, triggering a sharp neutrino burst that could be accompanied by a radio flare, probably caused by electron-positron acceleration at the same shock. 

Immediately following this sharp increase, relaxation processes within the jet sustain a slower and extended neutrino production at a lower intensity, which could be related to remnant radio emission from the jet relaxation. Even if proper simulations are needed to better characterize the neutrino timescale here, the jet relaxation –understood as the time needed to recover the conditions before the shock passage– may persist over approximately a timescale of $\sim R_{\rm jet}/c$ in the co-moving frame \citep{Fichet_2021, Fichet_2022}, gradually reducing the relativistic proton density. We can then assume that the neutrino flare stands for a time within $R_{\rm s} / c~<~t^\prime_\nu < R_{\rm jet} / c$, after which the proton density decreases. Although the star remains embedded within the jet for several years, significant neutrino production is largely confined to the intense month-to-year-long period (considering $R_{\rm jet} \sim 1~\rm{pc}$ and $\delta \sim 10$) immediately following the passage of the shock through the RG photospheric bubble. It is worth mentioning that other mechanisms could also lead to baryonic loading, such as contamination from a proton-rich sheath surrounding the jet. In fact, shear acceleration at the jet boundary could energize protons, as the velocity gradient between the fast inner jet spine and the slower surrounding medium provides a natural site for particle acceleration \citep{Rieger_2019}. However, its efficiency in producing localized and transient proton acceleration is limited, as this process is typically characterized by long acceleration timescales. This process is unlikely to be responsible for producing transient high-energy neutrino emissions, but could contribute to a quiescent steady state emission. \\

From the threshold energy production in $p\gamma$ interaction, it is possible to link the neutrino energy to the target photon field temperature (assuming thermal emission\footnote{We consider a thermal photon field in the lab frame, and not in the jet frame, which leads to a different relation than that in \cite{Ai_2023}, for example.}) as follows
 \begin{equation}
     E_\nu \approx \dfrac{0.05 \times 0.16~\rm{GeV}^2}{2.7~k_{\rm b} T_\gamma \left(1 + z\right)} \sim \left(\dfrac{3.5 \times 10^4}{T_\gamma \left(1 + z\right)}\right)~\rm{PeV} \,.
 \end{equation}
 \label{eq: Enu_T}

\begin{figure}
    \centering
    \includegraphics[width=\linewidth]{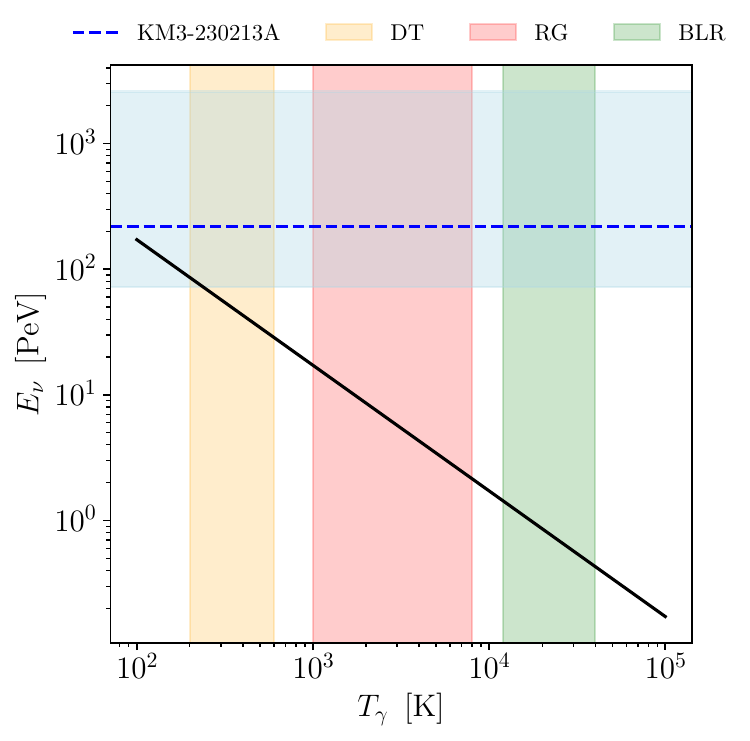}
    \caption{Expected peak neutrino energy $E_\nu$ from $p\gamma$ interactions as a function of the temperature $T_\gamma$ of the thermal target photon field (see Eq.~\ref{eq: Enu_T}). The dashed blue line displays the estimated energy for the neutrino event KM3$-$210213A within its $90\%$ confidence area (blue shade), and the dashed vertical lines represent typical ranges of temperatures of external photon fields (dusty torus (DT), red giant (RG), and broad line region (BLR)).}
    \label{fig: Enu_T}
\end{figure}

Figure~\ref{fig: Enu_T} shows the evolution of the neutrino energy as a function of the temperature of the target photon field, for a source at $z = 1$. For a typical RG star, we could expect neutrino energy ranging from $1-10~\rm{PeV}$, the predicted peak neutrino energy. As seen in Fig.~\ref{fig: Enu_T}, the KM3$-$230213A event could require a lower temperature of the photon field, similar to that expected from the dusty torus, in agreement with what is discussed in \cite{Neronov_2025}. A deeper study is needed to assess the feasibility of this alternative, but the physical framework developed here remains applicable, as the same jet–star interaction mechanism naturally provides the conditions for proton acceleration and subsequent $p\gamma$ interactions, regardless of the precise photon field responsible for the observed neutrino energy. \\

Although the shocked and thus dense stellar material could in principle drive local proton-proton {($pp$)} collisions -and yield a broad neutrino spectrum extending from tens of GeV up to PeV energies \citep[e.g.][]{Kelner_2006}, the diluted proton density is not high enough to trigger efficient neutrino production: For a typical red giant mass-loss rate of $10^{-8}~M_\odot \cdot \rm{yr}^{-1}$ \citep{Reimers_1977}, the expected proton density at the surface is 
\begin{equation}
    n_\star = n\left(r = R_\star\right) = \dfrac{\dot{M}}{4 \pi R^2_\star v_{\rm w} m_{\rm p}} \sim 3 \times 10^{8}~\rm{cm}^{-3} \,,
\end{equation}
using $R_\star = 10^{13}~\rm{cm}$ as the radius of the RG \citep{Piau_2011} and $v_{\rm w} = 10^{6}~\rm{cm}\cdot\rm{s}^{-1}$ for the wind speed \citep{Suzuki_2007}. At the stagnation point, the proton density becomes
\begin{equation}
    n_{\rm s} = n\left(r = R_{\rm s}\right) = n_\star \left(R_\star / R_{\rm s}\right)^2 \sim 10^{4}~\rm{cm}^{-3}\,.
\end{equation}
This value is much lower than the proton density required for the efficient $pp$ production of neutrinos. Only the proton density on the RG surface could be sufficient for $pp$ interactions. But even in that case, the interaction time ($\Delta t^\prime \sim R_\star / c \sim 10^4~\rm{s}$) would be too short. \\

\subsection{Detectability}
A remaining question is the capacity of current neutrino instruments to catch such a flare —and how often one might occur in the sky. IceCube, with $\sim15~\rm{yr}$ of full‐array operation, has recorded zero $>10~\rm{PeV}$ events, in contrast to the single KM3$-$230213A seen by KM3NeT in just a few years. \cite{Li_2025} showed that under standard diffuse or steady‐source assumptions, IceCube should have detected $\mathcal{O}\left(10\right)$ events above $10~\rm{PeV}$ —yet, none were found at $3-4~\sigma$ tension—, while allowing for a transient {event} reduces the expectation to $\ll 1$ event over IceCube’s lifetime and $\sim 1$ event in KM3NeT’s shorter exposure \citep{Robertson_2019, Li_2025}. In our scenario, each neutrino flare is tied to the brief timescale needed by the moving shock to cross the {RG bubble}, followed by relaxation. IceCube’s very small effective area for down-going cascades above $10^{17}~\rm{eV}$ naturally explains its non-detection of KM3$-$230213A, whereas KM3NeT’s larger acceptance for the same geometry makes a single $220~\rm{PeV}$ detection entirely plausible \citep{KM3NET_2016}.
Coordinated searches between IceCube and KM3NeT—and eventually next-generation telescopes—will be required to test this picture.\\

The previous paragraph collects evidence pointing towards neutrino bursts being caused by transient scenarios. One possibility would be to relate them to the propagation of perturbations through jets, as we have discussed before. At the forming region, a BZ jet is expected to be composed of pairs \citep{Blandford_1977, Parfrey_2019}. Different processes have been proposed to explain the launching of perturbations, such as magnetic reconnection or magneto-rotational turbulence \citep{Pino_2010, Lesur_2013}. Current VLBI monitoring shows an ejection rate of such radio components into an AGN jet of $\tau_{\rm ejecta} \sim 10^{-1}~\rm{yr}^{-1}$ \citep{Lister_2021}, which could be identified with travelling shocks \citep[e.g.,][]{Marscher_1985, Daly_1988, Gomez_1997, Agudo_2001,2009ApJ...696.1142M,Fromm_2016,Fichet_2021,Fichet_2022,2025A&A...693A.169S}. If the region around the RG is already above equipartition because of the proton mass-loading, when such an ejecta overruns the red-giant, it compresses the stellar bubble and increases the local baryon density. The interaction zone thus becomes strongly proton-dominated, $u_{\rm p}/u_{B} \gg 1$. In a similar interpretation, \cite{Kim_2025} pointed out a possible correlation between the recently high-energy neutrino event IceCube$-$211208A (IC211208) and a perturbation crossing a standing shock in the source PKS\,0735$+$178. \\

Regarding the rate of occurrence of these events, on the one hand, the RG number density in radio galaxies is poorly known, but \cite{Vieyro_2017} estimated the RG number density in M\,87's jet to be $n_\star \sim 10^{-2}~\rm{pc}^{-3}$. If we assume such a number density for our calculations, it results in a rather small number of RG within the first parsecs along the jet. On the other hand, {the CRATES catalog (in which PMN\,J0606$-$0724 is listed) contains $11131$ radio sources which show $S_{\rm 8.4} > 65~\rm{mJy}$ \citep{Healey_2007}. As the flux criterion selects Doppler-boosted cores, the sample is already strongly biased toward small viewing angles. We therefore adopt $N_{\rm radio} \sim 1.1 \times 10^{4}$ as the population of aligned, radio-loud jets relevant in our scenario. In this context, $N_{\rm radio}$ should be considered like an upper limit as it might contain sources with too high viewing angle, or too high redshift, but overall as a representative number of the sources of interest. Estimating the rate of shock-star interactions within a jet volume of $V_{\rm jet} \sim 1~\rm{pc}^3$ leads to
\begin{equation}
    \dot{N}_{\rm RG-shock}  = N_{\rm radio}~ n_\star  ~\tau_{\rm ejecta}~V_{\rm jet} \lesssim 10~\rm{yr}^{-1} \,.
\end{equation}
Hence, we can expect $\lesssim 100$ red‐giant–shock encounters over the lifetime of current $\rm{km}^3$ neutrino detectors. As previously mentioned, this number appears as an upper limit, as not every encounter could produce a detectable neutrino flux, and detailed simulations (varying jet parameters, moving shock characteristics, RG number density, etc.) are required to assess the true rate of shock-star interactions leading to the production of detectable neutrino events.} \\

{Beyond individual detections, it is also instructive to assess whether red-giant shock interactions could contribute appreciably to the diffuse neutrino background observed by a neutrino instrument as IceCube. In that context, we can estimate the average per-source neutrino flux expected from a population of radio galaxies, and compare it to detector sensitivities. We can therefore consider the full} diffusive neutrino flux seen by the IceCube detector \citep[all neutrino flavors, on all the sky; e.g., ][]{Naab_2023}, 

\begin{equation}
    E_{\rm \nu, tot} \Phi_{\rm \nu , tot}\left(E_\nu\right) = 4.2 \times 10^{-14}~E^2_\nu \left(\dfrac{E_\nu}{100~\rm{TeV}}\right)^{-2.52}~\rm{erg}\cdot\rm{cm}^{-2}\cdot\rm{s}^{-1} \,.
\end{equation}

If neutrinos are equally emitted from $N_{\rm radio}$ sources, we can get an average of a maximal neutrino flux per source at $1~\rm{PeV}$ of
\begin{equation}
    E_{\rm \nu, radio} \Phi_{\rm \nu, radio} \left(E_\nu = 1~\rm{PeV}\right) \leq 1.1 \times 10^{-14}~\rm{erg}\cdot\rm{cm}^{-2}\cdot\rm{s}^{-1} \,,
\end{equation}
which should go down at $\sim 1.1\times 10^{-15}~\rm{erg}\cdot\rm{cm}^{-2}\cdot\rm{s}^{-1}$ assuming that $10\%$ of the extragalactic neutrino diffuse flux, at $1~\rm{PeV}$, is coming from blazars. {In our context, this means} that on average, quiescent neutrino emissions from RG in the jet are below detectors' sensitivities. Following our scenario, the signal might be detectable above background for either brief shock-star interactions, which could bring the neutrino flux close to the IceCube / KM3NeT detection limit, or over years in the mean of stacking analysis (as we described above).

\subsection{Observational predictions and application to PMN\,J0606$-$0724}

Building upon the scenario described above, in which a RG star embedded in a jet interacts with a travelling perturbation within a lepton-dominated jet, we now outline the key observational signatures predicted by this mechanism and compare them with the available data for PMN\,J0606$-$0724. First, our model requires
\begin{itemize}
\item Blazar-like objects whose jets are closely aligned with our line of sight, so that Doppler boosting maximizes the neutrino flux that reaches the Earth.
\item A moving shock (a radio “knot’’) observed and accompanied by a radio flare.
\item A quiescent neutrino flux punctuated by a local neutrino outburst when the shock encounters a red-giant envelope; the outburst is expected to occur during the radio flare, its duration may reflect the time it takes for the moving shock to traverse the stellar envelope and reveal the position of the RG bubble in the jet. This connection between the neutrino flare and shock propagation timescale has been recently modeled for PKS\,0735$+$178 by \citet{Kim_2025}.
\item A dense, thermal photon field—produced and sustained, in our case, by the RG atmosphere crossing the jet—which could be spatially resolved in nearby sources \citep{2014A&A...569A.115M}.
\item Suppressed or absent GeV–TeV $\gamma$-ray emission. In our scenario, neutral pion decays would inject a $\gamma$-ray luminosity comparable to the neutrino one, but these photons are efficiently absorbed through $\gamma\gamma$ pair production against the boosted thermal field/jet photon field. The resulting $e^\pm$ pairs cool via synchrotron radiation, reprocessing the energy into the MeV-GeV band, where the predicted flux at $z\simeq1.2$ falls below current instrument sensitivities. Additional attenuation on the EBL further suppresses any residual TeV signal along the line of sight.
\end{itemize}

Some of these predictions align well with the observational data from PMN\,J0606$-$0724 during the ejection event. The absence of a strong GeV-TeV signal could be explained by both internal absorption and {EBL} attenuation at the source redshift $z=1.227$. At this redshift, the $\gamma$-ray flux cutoff is expected around $>40~\rm{GeV}$ \citep{Dominguez_2011}. Furthermore, the estimated radio power $10^{43}\rm{erg}\cdot\rm{s}^{-1}$ from the observed radio flux at $15~\rm{GHz}$ and the \textit{Fermi}-LAT $\gamma$-ray upper limit of $<10^{46}\rm{erg}\cdot\rm{s}^{-1}$ are both compatible with a BL~Lac object with a jet power, around $10^{45-46}\rm{erg}\cdot\rm{s}^{-1}$. However, at the time of the writing, no available VLBI observations permit us to conclude on the presence of moving knots in the jet of PMN\,J0606$-$0724. Regarding the source of thermal photons, our results indicate that the red-giant photosphere ($T_\gamma \sim 3$–$5\times10^3$ K) produces neutrinos in the $1-10~\rm{PeV}$ range (see Fig.~\ref{fig: Enu_T}), consistent with expectations from $p\gamma$ interactions on UV photons. Neutrinos with $E_\nu \gtrsim 100~\rm{PeV}$, as in KM3$-$230213A, are instead more naturally explained by interactions with colder infrared photons from the dusty torus. In our scenario, the red giant supplies the baryons, the moving shock provides the acceleration, and the encounter rate sets the overall probability, while the dominant photon field determines the neutrino energy scale. 

As we show, such events are expected to be rare as over the operational lifetime of current $\rm{km}^3$ neutrino detectors, only a tenth of such interactions were expected to be detectable in the past $15$ years. As pointed out by \cite{Neronov_2025} in the case of KM3$-$230213A, a hard proton spectrum could also explain why lower energetic neutrinos have not been observed. In our case, we consider a simple Fermi-type acceleration, but reacceleration at shocks and more complex mechanisms such as stochastic acceleration could be expected and explain such a spectrum \citep{Liu_2017}. \\

In conclusion, PMN\,J0606$-$0724 exhibits some characteristics expected from the jet-RG interaction scenario outlined here. While not every such encounter will produce observable neutrino signals, this framework offers a potential explanation for the production of high-energy cosmic rays. 

\section{Conclusions}
\label{sec: conclusion}

Using the measured energy and flux of KM3$-$230213A, we infer a required proton luminosity of about $10^{47}~\rm{erg}\cdot\rm{s}^{-1}$. Our analysis shows that this power is accompanied by a strong departure from equipartition, with the proton energy density exceeding the magnetic one by several orders of magnitude. Among the seventeen blazars located within the $3^\circ$ angular error region, PMN\,J0606$-$0724 is the only source showing a radio flare at $15~\rm{GHz}$ peaking within days of the neutrino detection, and whose long-duration light curve aligns with the high-power, month-to-year-long transient required by our energetics. This makes it the most plausible electromagnetic counterpart to KM3$-$230213A. 

We propose a scenario in which a fast internal shock, launched from the jet base, collides with a red giant star embedded in the jet at parsec scales \citep[see also, e.g.,][]{Barkov_2010,2012ApJ...749..119B,Bosch_2012,2013ApJ...774..113K}. The resulting compression of the stellar envelope briefly increases the local proton number density and boosts the photon field, enabling efficient photohadronic interactions and generating a PeV neutrino burst on the month-to-year scale due to the jet relaxation \citep[$\propto R_{\rm jet} / c \delta$, e.g.,][]{Fichet_2022}. The same event can account for the observed radio flare, while the absence of a coincident GeV–TeV signal is consistent with the expected $\pi^0$ decay channel: the associated $\gamma$-ray luminosity, comparable to the neutrino one, would be efficiently absorbed through $\gamma\gamma$ interactions with the boosted stellar photon field. The resulting $e^\pm$ pairs cool rapidly via synchrotron emission, reprocessing the energy into the MeV-GeV band. At the redshift of PMN\,J0606$-$0724, the predicted flux falls below the sensitivity of \textit{Fermi}-LAT, explaining the lack of a detectable high-energy counterpart.

In this picture, the peak neutrino energy is set by the temperature of the target photon field. For a red giant with a surface temperature of a few thousand Kelvin, boosted into the jet frame, photon energies enter the ultraviolet range, enabling PeV-scale neutrino production. 

Overall, our model —centered on localized, transient proton loading triggered by interaction with external material— offers a robust physical framework for producing PeV neutrinos in AGN jets. While we discussed it here in the specific case of KM3$-$230213A, the scenario can be generally applied. Future neutrino telescopes such as IceCube-Gen2 and KM3NeT Phase-2 may detect additional high-energy events where similar conditions apply. Our results suggest that radio-bright AGN, particularly those exhibiting recurrent moving radio components interpreted as shocks, in combination with stars crossing the jet, could be prime candidates for neutrino production. Systematic searches targeting this subclass —especially high-excitation blazars– may significantly enhance the chances of identifying neutrino counterparts and refining our understanding of the origins of the high-energy cosmic neutrino flux.

\section*{acknowledgements}
This work has been supported by the Spanish Ministry of Science through Grant \texttt{PID2022-136828NB-C43}, from the Generalitat Valenciana through grant \texttt{CIPROM/2022/49}, from the Astrophysics and High Energy Physics program supported by the Spanish Ministry of Science and Generalitat Valenciana with funding from European Union NextGenerationEU (\texttt{PRTR-C17.I1}) through grant \texttt{ASFAE/2022/005}.

\section*{Data availability}
The data underlying this article are available in the article and in its online supplementary material.

\bibliographystyle{mnras}
\bibliography{bib}

\bsp	
\label{lastpage}
\end{document}